\documentclass[twocolumn,showpacs,preprintnumbers,amsmath,amssymb]{revtex4}
\usepackage{latexsym}
\usepackage{graphicx}
\usepackage{dcolumn}
\usepackage{bm}
\usepackage{xcolor}

\input amssym.def
\input amssym.tex

\newcommand{\half}{{{\textstyle\frac{1}{2}}}}

\newcommand{\be}{\begin{equation} }
\newcommand{\ee}{\end{equation} }
\newcommand{\beqa}{\begin{eqnarray} }
\newcommand{\eeqa}{\end{eqnarray} }
\newcommand{\ba}{\begin{array}}
\newcommand{\ea}{\end{array}}

\newcommand{\so}{\mathbf{so}}

\newcommand\tr{{\rm tr}}
\newcommand\Tr{{\rm Tr}}
\newcommand\const{{\nu}}
\newcommand\rmd{{\rm d}}
\newcommand\rmx{{\rm x}}
\newcommand\bfx{{\bf x}}

\newcommand\cG{{\cal G}}
\newcommand\cH{{\cal H}}

\newcommand\cL{{\cal L}}
\newcommand\cM{{\cal M}}

\newcommand\cS{{\cal S}}
\newcommand\cT{{\cal T}}

\newcommand\halpha{{\hat{\alpha}}}
\newcommand\hbeta{{\hat{\beta}}}

\newcommand\matter{{\rm \scriptscriptstyle{}}}

\newcommand\PB{{\rm \scriptscriptstyle{PB}}}
\newcommand\NB{{\rm \scriptscriptstyle{NB}}}

\newcommand\Poly{{{\rm Poly}}}

\newcommand\CFT{{{\rm CFT}}}

\newcommand\Mt{{M(t)}}
\newcommand\mD{m}
\newcommand\mnut{{{m}_{\nu}}(t)}

\newcommand\SF{{{{a}}}}
\newcommand\rX{{r(\rho)}}
\newcommand\rtwoX{{r^{2}(\rho)}}

\newcommand\dis{\displaystyle}

\newcommand\Mtwo{\scriptstyle{\rm{M2}}}

\begin{document}
\preprint{MPP-2007-5}
\title{Spacetime Emergence in the Robertson-Walker Universe from a Matrix model}

\author{Johanna Erdmenger$^{\ast}$, Ren\'{e} Meyer}
 \altaffiliation{
 Max-Planck-Institut f\"{u}r Physik,  
 80805  M\"{u}nchen, Germany.}
\author{Jeong-Hyuck Park}
 \altaffiliation{
 Dept. of Physics, Sogang University, Seoul 121-742, Korea.}

\date{\today}

\begin{abstract}
Using a novel,  string theory-inspired
formalism based on a Hamiltonian constraint, we obtain
a conformal mechanical system for the spatially flat
four-dimensional Robertson-Walker Universe. Depending on parameter
choices, this system  describes either a relativistic particle in
the Robertson-Walker background, or metric fluctuations of the
Robertson-Walker geometry. Moreover we derive a tree-level
$\cM$-theory   matrix model
in this time-dependent background.
Imposing the Hamiltonian constraint  forces the
spacetime geometry to be fuzzy  near the big bang, while the
classical Robertson-Walker geometry emerges  as the Universe
expands. From our  approach we also derive
the temperature of the Universe interpolating between the radiation and
matter dominated eras.
\end{abstract}
\pacs{98.80.-k, 11.25.Yb}
\maketitle

Recent astronomical data show that the current Universe is very close to
the spatially flat  Robertson-Walker (RW) geometry\,\cite{Spergel:2006hy}.
The Universe has evolved from a big bang singularity, near which
quantum effects are expected to have played an important role.
While a complete quantum gravity description of the big bang is
unavailable, effective matrix model descriptions of string/$\cM$-theory on time
dependent backgrounds have lead to a number of insights
\cite{HashiSethi,Craps}.
However, the main focus of these studies so far has been on
time-dependent and  supersymmetry-preserving
orbifold or plane wave backgrounds,   but not on the
 physically relevant supersymmetry-breaking  RW geometry.
One technical obstacle is that the latter lacks a null
isometry. Hence the conventional light-cone quantization is not applicable
and a new approach is required. We  develop such an approach in the present
Letter:~the characteristic feature of our formalism is the presence of  a
\textit{Hamiltonian constraint},~\textit{i.e.}  a vanishing energy constraint.
Instead of  fixing  the light-cone momentum,
we consider a sector of fixed Hamiltonian density.
In this way, for the first time it becomes possible, at least at tree level,
to construct an M2-brane or $\cM$-theory  matrix model\,\cite{Banks:1996vh}  for
the realistic  RW geometry
and demonstrate   the emergence of classical spacetime from an
originally fuzzy geometry.\newline
\indent In this Letter we begin  with the analysis of the geodesic motion of a
\textit{single} D-particle in the  RW Universe. In particular,  we propose
a conformal mechanics model invariant under
one-dimensional diffeomorphisms.
For two different parameter
regimes and gauge choices, this mechanical system describes \textit{either}
the geodesic motion
of a  point particle in the spatially flat RW background, \textit{or}
homogeneous metric fluctuations around the background.
More precisely, in each case we  find a
conserved quantity and  show that any  sector of the fixed value of
the quantity is described by the conformal mechanics.
This is reminiscent of the AdS/CFT correspondence\,\cite{Maldacena:1997re},
where matter and gravity dynamics are mapped to each other.
Here however both the descriptions of matter and gravity descend from the same
CFT action. Finally, we
derive a  matrix model from the action of  the bosonic M2-brane
in the RW background, giving
a many-particle gene\-ralization of the single-particle action,
as in flat space\,\cite{Banks:1996vh}.
We show that imposing the Hamiltonian constraint in the matrix model
ensures the emergence of spacetime.
Emergence here means that the Hermitian matrices in the matrix
model, whose eigenvalues encode the positions of D-particles, become
simultaneously diagonalizable far away from singularities, such that
the particle positions can be simultaneously
measured, and geometric quantities become classical\,\cite{Craps}.

\section{Conformal Mechanics   \label{genconfmech}}
In $D$-dimensional spacetime, requiring both homogeneity and isotropy
of the $D{-1}$ spatial dimensions, the metric is   constrained to the
RW geometry\,\cite{Weinberg}
\begin{equation}
\displaystyle{\rmd s^{2}=-\rmd t^{2}+\SF^{2}(t)\!\left[\rmd\bfx^{2}+
\kappa ({\bfx\cdot\rmd\bfx})^{2}\!/(1-\kappa\bfx^{2})\right]\,,}
\label{metric0}
\end{equation}
where $\kappa$ is a constant, 
and  $\SF(t)$ is the only undetermined  scale factor
depending on the cosmic time $t$.\newline
\indent The conformal mechanics (``CFT"),
 which we  will show to be  closely  related  to the spatially
flat ($\kappa{=0}$)  RW Universe, is of  the  general form
\begin{equation}
\dis{\cS_{\CFT}=\int\!\rmd t\left[\half\eta\dot{\varphi}^{2}+\half\eta^{-1}
\left(\frac{\,c_{1}}{\,\varphi^{2}}+c_{2}\right)\right]}\,.
\label{LCFT}
\end{equation}
Here $c_{1}$, $c_{2}$ are constants, $\varphi(t)$ is the only
dynamical variable and $\eta(t)$ is the inverse of an
einbein. Both $\varphi$ and $\eta$ transform under one-dimensional
diffeomorphism  $t\rightarrow s(t)$ as
$(\varphi(t),\eta(t))\rightarrow(\varphi(s),\eta(s)/\dot{s})$.

Integrating out the auxiliary variable $\eta$ reduces the action to
$\int\!\rmd\varphi\sqrt{c_{1}\varphi^{-2}+c_{2}}$. On the other hand,
fixing the gauge symmetry with an arbitrary function of time,
$\eta\equiv\hat{\eta}(t)$, the mechanical system~(\ref{LCFT}) essentially
corresponds to the known conformal mechanics\,\cite{deAlfaro:1976je}.
The gauge fixed action is invariant under the  transformation
$\delta\varphi=\hat{\eta}(f\dot{\varphi}-\half \dot{f}\varphi)$,
where $f(t)$
is given by a solution of $\frac{\rmd~}{\rmd t}[\hat{\eta}\frac{\rmd~}{\rmd t}
\!(\hat{\eta}\dot{f})]=0$.
This third order differential equation has three solutions  which
form  the symmetry algebra $\so(1,2)$\,\footnote{
With $\beta(t):=\!\int\!\rmd t\,(l\hat{\eta}(t))^{-1}$,
$p_{\varphi}{=\hat{\eta}\dot{\varphi}}$,
three solutions are
${f}_{0}={\textstyle{\frac{1}{\sqrt{2}}}}
\left(l^{2}+\half\beta^{2}\right)$, ${f}_{1}={l\beta}$,
${f}_{2}={\textstyle{\frac{1}{\sqrt{2}}}}
\left(l^{2}-\half\beta^{2}\right)$,  and
the corresponding Noether charges form $\so(1,2)$ Lie algebra,
$\{Q_{{1}},Q_{{2}}\}_{\PB}=-Q_{{0}}$, $\{Q_{{2}},Q_{{0}}\}_{\PB}
=Q_{{1}}$, $\{Q_{{0}},Q_{{1}}\}_{\PB}=Q_{{2}}$, where
$Q_{f}=\half \!f\!\left(p_{\varphi}^{2}-c_{1}\varphi^{-2}\right)
-\half\hat{\eta}\dot{f}\varphi
p_{\varphi}+\frac{1}{4}\hat{\eta}\varphi^{2}\frac{\rmd~}{\rmd t}
(\hat{\eta}\dot{f})$ satisfying
$\{\varphi,Q_{f}\}_{\PB}=\delta_{f}\varphi$.}.\newline
\indent The Hamiltonian corresponding to the gauge fixed Lagrangian is,
with the canonical momentum $p_{\varphi}=\hat{\eta}(t)\dot{\varphi}$,
\begin{equation}
\dis{\cH_{\CFT}=\frac{1}{\,2\hat{\eta}(t)}\left(p_{\varphi}^{2}-
\frac{\,c_{1}}{\,\varphi^{2}}-c_{2}\right)\,.}
\label{HCFT}
\end{equation}
The physical states must
lie on the surface of vanishing energy $\cH_{\CFT}\equiv 0$ in the phase space, as
implied by the gauge fixing of  diffeomorphism invariance.
(Note that throughout the Letter, `$\equiv$' denotes  gauge fixings
or on-shell relations.)


\subsection{CFT for Single D-particle Dynamics\label{CFTD}}
Our starting point is a novel formalism for
a generic mechanical system satisfying the following two
conditions: (i) The Hamiltonian is given by the inverse of the
Lagrangian
\begin{equation}
\cH\cL=-m^{2}\,,
\label{HLm}
\end{equation}
where $m$ is a constant mass parameter.
This always holds for a square-root
relativistic particle Lagrangian of the form
$\cL=-m\sqrt{1-g_{ij}(t,x)\dot{x}^i\dot{x}^j\,}$,
after the gauge fixing to identify the worldline affine parameter with
the cosmic time.~~ (ii) There exists a conserved quantity with on-shell
value $\nu$, such that  for the sector of fixed $\nu$ the Lagrangian is
completely fixed on-shell as a time- and $\nu$-dependent function,
\begin{equation}
\cL\equiv e_{\nu}(t)\,.
\label{Lonshellnu}
\end{equation}
Then \textit{the square-root free Lagrangian}
\begin{equation}
\displaystyle{\cL_{\nu}:=\frac{\cL^{2}}{\,2e_{\nu}(t)}-\frac{m^{2}}{\,e_{\nu}(t)}
+\frac{e_{\nu}(t)}{2}}
\label{geLnu}
\end{equation}
\textit{together with the Hamiltonian constraint
equivalently
describes the sector of fixed $\nu$}\,\footnote{\textit{cf.}~\eqref{geLnu}
and  Polyakov Lagrangian  $\cL_{\Poly}=\half\left( e^{-1}\cL^{2}+ e\right)$.}.
This can be shown by observing that all the
canonical momenta of~(\ref{geLnu}) take the same on-shell values as those of
$\cL$. Further, the  Hamiltonian corresponding to $\cL_{\nu}$  reads
from (\ref{HLm}) and (\ref{geLnu})
\begin{equation}
{\cH_{\nu}=\left(\cL^{2}-e_{\nu}^{2}\right)\!/(2e_{\nu}).}
\end{equation}
The Hamiltonian constraint\,$\cH_{\nu}\equiv 0$ then implies (\ref{Lonshellnu}).
Henceforth, as an example (see also \cite{Erdmenger:2006eh}),   we turn to
a relativistic particle in  four-dimensional RW
background\,(\ref{metric0}) in spherical coordinates
\begin{equation}
\displaystyle{\rmd s^{2}=-\rmd t^{2}+\SF^{2}(t)\Big[
\rmd \rho^{2}+\rtwoX\left(\rmd\theta^{2}+\sin^{2}\!\theta\,\rmd\phi^{2}\right)\Big]\,,}
\nonumber
\end{equation}
where ${\rX=\sqrt{\rmx^{2}}}\,$ is given by
$\sin\!\left(\sqrt{\kappa}\rho\right){/\sqrt{\kappa}}$,
$\rho$ or $\sinh\!\left(\sqrt{-\kappa}\rho\right){/\sqrt{-\kappa}}$
depending on $\kappa$ being positive, zero or negative, respectively.
After identifying the worldline affine parameter with cosmic
time, the point particle or D-particle\,\footnote{Any possible homogeneous
dilaton factor can be absorbed into the RW metric \textit{via} redefinition of `frame'.}
Lagrangian  in the RW background  becomes
\begin{equation}
\displaystyle{\!\!\!\!\!\cL_{\matter}=
-\mD\sqrt{1-\SF^{2}(t)\left(\dot{\rho}^{2}+\rtwoX
\dot{\theta}^{2}+\rtwoX\sin^{2}\!\theta\,\dot{\phi}^{2}\right)}}\,.
\label{relatL}
\end{equation}
The canonical momenta for  $\rho$, $\theta$ and $\phi$ are
\begin{equation}
\ba{lll}
p_{\rho}=\Mt\dot{\rho}\,,&\!
p_{\theta}=\Mt\rtwoX\dot{\theta}\,,&\!
p_{\phi}=\Mt\rtwoX\sin^{2}\!\theta\,\dot{\phi}\,,
\ea
\label{momenta}
\end{equation}
where we set  $M(t):=-{\mD^{2}\SF^{2}(t)}/{\cL_{\matter}}$ for compact expression.
The corresponding Hamiltonian
\begin{equation}
\displaystyle{\cH_{\matter}
=\sqrt{\mD^{2}+ \SF^{-2}(t)\left[p_{\rho}^{2}
+{r^{-2}(\rho)}\left(
p_{\theta}^{2}+{p_{\phi}^2}/{\,\sin^2\!\theta}\right)\right]}\,,}
\nonumber
\end{equation}
satisfies \,(\ref{HLm}).
In spite of the arbitrariness of $\SF(t)$, the
dynamics is \textit{integrable}, as there exist three mutually
Poisson-bracket commuting conserved quantities
\begin{align}\label{cons1}
p_{\phi} & \equiv  \mbox{constant}\,,\\\label{cons2}
J^{2} := p_{\theta}^{2}+p_{\phi}^{2}/\sin^{2}\!\theta &\equiv j(j+1)\,,\\
p_{\rho}^{2} +J^{2}/r^{2}(\rho) &\equiv (\const m)^{2}\,.\label{cons3}
\end{align}
The dimensionless constant $j$ plays the role of a classical $\so(3)$ angular momentum.
Introducing a time dependent mass as an on-shell value of $M(t)$,
\begin{equation}
\displaystyle{\mnut:=\mD\SF(t)\sqrt{\SF^{2}(t)+\const^{2}}\equiv M(t)\,,}
\label{mtdef}
\end{equation}
the Hamiltonian and  the Lagrangian assume the on-shell values,
$\cH_{\matter} = -m^{2}/\cL \equiv \mnut/a^{2}(t)$.
As for (\ref{geLnu}) we have
\begin{equation}
\displaystyle{\cL_{\nu}=\frac{\,\mnut}{2}
\!\left(\dot{\rho}^{2}+
\rtwoX\dot{\theta}^{2}+\rtwoX\sin^{2}\!\theta\,\dot{\phi}^{2}\right)+
\frac{\,\left(\const\mD\right)^{2}}{\,2\mnut}\,.}
\label{nonrelLG}
\end{equation}
Again, all the off-shell canonical momenta of (\ref{nonrelLG})
match with  the on-shell ones of (\ref{relatL}).
Further, the corresponding Hamiltonian
\begin{equation}
\displaystyle{\cH_{\nu}=\frac{1}{\,2\mnut}
\left(p_{\rho}^{2}+\frac{J^{2}}{\,\rtwoX}-\left(\const\mD\right)^{2}\right)}
\label{Hnew}
\end{equation}
exhibits the same mutually commuting conserved quantities as
(\ref{cons1})-(\ref{cons3}).
Thus the surface of the vanishing energy $\cH_{\nu}\equiv 0$ in
the phase space of the dynamical system\,(\ref{Hnew})
describes precisely
the relativistic particle
in the RW background for a sector of fixed $\const$. Further  the
subsector of fixed angular momentum is reached by setting
$J^2\!\equiv\!j(j{+1})$, which reduces   (\ref{Hnew})
to the conformal mechanics\,\eqref{HCFT}.   In this way,
\textit{the conformal mechanics\,\eqref{HCFT}
with the  choice $\varphi{=\rho}$,
$\hat{\eta}{=\mnut}$, $c_{1}=\!-j(j{+1})\leq 0$, $c_{2}=(\nu m)^{2}\geq0$
describes     the geodesic motion of a
relativistic particle with respect to  cosmic time
in the spatially flat RW Universe, with fixed conserved
quantities $\nu$, $j$.}
\subsection{CFT for Homogeneous Gravity\label{CFTGR}}
Pioline and Waldron\,\cite{Pioline:2002qz} observed that for a solely time-dependent,
\textit{generic} $D$-dimensional metric in longitudinal gauge
\begin{equation}
\ba{ll}
\displaystyle{\!\!\!\!\!
\rmd s^{2}=-e^{2}\varrho^{-2}\rmd t^{2}+\varrho^{{2}/{(D-1)}}
\hat{g}_{ij}\rmd x^{i}\rmd x^{j}\,,}\,&\det\hat{g}=1\,,
\ea
\label{homogeneousmetric}
\end{equation}
the Einstein-Hilbert action with cosmological constant reduces to a mechanical system
 for a relativistic  ``fictitious" point particle,
\begin{equation}
\displaystyle{
\!\!-\!\!\int\!\rmd^{D}\!x\sqrt{-{g}}\left(R{-2\Lambda}\right)\!=\!V\!\!\!
\int\!\rmd t\,
\frac{1}{e}\!\left(\!\frac{\!\,D{-2}}{\!\,D{-1}}\!\right)\dot{\varrho}^{2}
-2e\!\left(\!\frac{\hat{C}}{\,\varrho^{2}}-\Lambda\!\right).}
\nonumber
\end{equation}
Here $V$ is  the $D{-1}$ dimensional spatial volume, henceforth normalized
to one, and $\hat{C}:=\frac{1}{8}e^{-2}\varrho^{4}\tr(\hat{g}^{-1}\dot{\hat{g}}
\hat{g}^{-1}\dot{\hat{g}})$ which contains  a non-linear $\sigma$-model metric for
the  coset ${{\mbox{SL}}{(D}-1)}/{{\mbox{SO}}{(D}-1)}$ as
$\tr(\hat{g}^{-1}\dot{\hat{g}}
\hat{g}^{-1}\dot{\hat{g}})=h_{ab}(\theta)\dot{\theta}^{a}\dot{\theta}^{b}$.
In  terms of the
momenta\,$p_{\varrho}=\omega\dot{\varrho}/(4e)$,
$p_{a}=-\half\varrho^{2}e^{-1}h_{ab}\dot{\theta}^{b}$, we identify
$\hat{C}=\half h^{ab}(\theta)p_{a}p_{b}$ as the kinetic energy on the coset space
and write the Hamiltonian
\begin{equation}
\ba{ll}
\!\!\cH_{\mathrm{E.H.}}=\displaystyle{\frac{2e}{\omega}\!
\left(p_{\varrho}^{2}-\frac{\,\omega\hat{C}}{\,\varrho^{2}}
-\omega\Lambda\right)}\,,&\dis{\omega:=8\!\left(\frac{D{-2}}{D{-1}}\right)\,.}
\ea
\label{HEH}
\end{equation}
$\hat{C}$ is conserved\,\cite{Pioline:2002qz} and, in fact, positive
semi-definite\,\footnote{With the diagonalization
$\hat{g}=o\lambda o^{t}$, $oo^{t}=1$, $\lambda>0$,
$\tr\!\left(\hat{g}^{-1}\dot{\hat{g}}
\hat{g}^{-1}\dot{\hat{g}}\right)=2
\sum_{i>j}\left[(\!\sqrt{\lambda_{i}{/\lambda_{j}}}-\sqrt{\lambda_{j}{/\lambda_{i}}}\,)
(o^{t}\dot{o})_{ij}\right]^{2}+\sum_{i}(\dot{\lambda}_{i}/\lambda_{i})^{2}\geq 0$.}.

%
%

The original motivation of \cite{Pioline:2002qz} to consider the homogeneous
modes only was based on the observation in\,\cite{Belinsky} that near
cosmological  singularities,   inhomogeneous modes generically decouple.
Here we make the further   observation that
the spatially flat RW geometry\,\eqref{metric0}
is a special case of \eqref{homogeneousmetric} as
$e=\varrho=a^{D-1}$, $\hat{g}_{ij}=\delta_{ij}$, and hence \eqref{homogeneousmetric} is
the most general homogeneous and non-perturbative  fluctuation of the RW metric.
In the cosmic time gauge~$e=\varrho$,
\textit{the conformal mechanics\,(\ref{HCFT}) with the choice $\varphi{=\varrho}$,
$\hat{\eta}{=\frac{1}{4}\omega a^{1{-D}}}$ or $\eta=\omega/(4\varphi)$,
$c_{1}\equiv\omega\hat{C}\geq 0$,
$c_{2}=\omega\Lambda$  describes the homogeneous metric fluctuations of
the spatially flat RW Universe  with respect
to  cosmic time,
with  fixed  value of  kinetic energy on the coset space.}
In particular,  near the big bang ($a{\simeq 0}$), the choice of small
$\hat{\eta}=ma\sqrt{a^{2}+\nu^{2}}$ and negative $c_{1}$
describes the matter, whilst large
$\hat{\eta}{=\frac{1}{4}\omega a^{1{-D}}}$ and positive $c_{1}$ describes the gravity.
As an example for the map between gravity and matter,
the metric element  $\varrho=\sqrt{\det g_{ij}}$ in gravity is mapped to
the radial coordinate $\rho$ of the particle trajectory.

\section{Matrix model for RW Universe\label{MMM}}
Now we turn to the description of \textit{many} D-particles in the spatially flat RW
Universe. The description  is generically given by a
Yang-Mills quantum mechanics\,\cite{Witten:1995im},~\textit{i.e.}~by a matrix model
generalization of a single particle action.  In a flat background, the
coupling of the Yang-Mills potential $[X^{i},X^{j}]^{2}$
in the matrix model
can be freely scaled and therefore  is irrelevant. However, in
the RW background the coupling coefficient is
 time-dependent, and cannot be simply  deduced
from the one-particle action\,(\ref{nonrelLG}).
We determine this  time-dependent coupling
by deriving the  tree-level $\cM$-theory matrix model
from the bosonic M2-brane action in the RW background.
The dynamics of an M2-brane with tension $T$ embedded in a
$D$-dimensional target spacetime is governed by the
Nambu-Goto action
\begin{equation}
S_{\mathrm{M2}}=-T\int \rmd^3 \xi \sqrt{-\det\!\left(
\partial_{\halpha}x^{\mu}\partial_{\hbeta}
x^{\nu}g_{\mu\nu}(x)\right)}\,.\label{NambuGoto}
\end{equation}
The three-dimensional worldvolume of the M2-brane
is parameterized by coordinates $\xi^{\halpha}$,
 $\halpha=0,1,2$, and the embedding functions are given by $x^{\mu}(\xi)$,
 $\mu=0,1,\cdots,D{-1}$.  Adopting the cosmic gauge $t=x^{0}=\xi^{0}$ as well as
the longitudinal gauge $\partial_{t}x^{\mu}\partial_{\alpha}
x^{\nu}g_{\mu\nu}=0$, $\alpha=1,2$ (see \cite{Kim:2006wg} for a
detailed procedure), the remaining worldvolume diffeomorphism is, at this stage,
static,~\textit{i.e.}~$\xi^0$-independent. The Nambu-Goto Lagrangian in the
spatially flat RW background then reduces to
\begin{equation}
\cL_{\Mtwo}=-Ta^{2}(t)
\sqrt{\left(1-a^{2}(t)\dot{x}^{2}\right)
\det\cG}\,,
\label{LM2}
\end{equation}
where the determinant
$\det \cG=\det (\partial_{\alpha}x^{i}\partial_{\beta} x_{i})$,  $\alpha,\beta=1,2$,
is taken over spatial M2-brane coordinates only.
Spatial indices $i,j$ are contracted with the flat metric $\delta_{ij}$ such that
$\dot{x}^2{:=\dot{x}^{i}\dot{x}^{j}\delta_{ij}}$.
With the momenta
$p_{i}=Ta^{4}\dot{x}_{i}\sqrt{{\det\cG}/({1-a^{2}\dot{x}^{2}})}$,
the equation of motion reads
\begin{equation}
\dot{p}_{i}\equiv\partial_{\alpha}\left[Ta^{2}\!
\sqrt{\left(1-a^{2}\dot{x}^{2}\right)\det\cG}\,
\cG^{-1\alpha\beta}\partial_{\beta}x_{i}\right]\,,
\label{M2EOM}
\end{equation}
and the longitudinal gauge condition becomes
\begin{equation}
\ba{lll}
p_{i}\partial_{\alpha}x^{i}=0~~~\Longleftrightarrow~~~
\dot{x}_{i}\partial_{\alpha}x^{i}=0\,.
\ea
\label{longiM2}
\end{equation}
In terms of the Nambu bracket\, $\left\{x,y\right\}_{\NB}\!:
=\epsilon^{\alpha\beta}\partial_{\alpha}x
\partial_{\beta}y$
($\epsilon^{12}=1$, $\epsilon^{\alpha\beta}=
-\epsilon^{\beta\alpha}$),
the determinant can be expressed as
$\det\cG=\half\left\{x^{i},x^{j}\right\}_{\NB}\!\left\{x_{i},x_{j}\right\}_{\NB}$.
Further  (\ref{M2EOM}) and (\ref{longiM2}) imply respectively
\begin{equation}
\ba{ll}
\partial_{t}\!\left(p^{2}\right)+
T^{2}a^{6\,}\partial_{t}\!\det\cG\equiv 0\,,
~&~\left\{p_{i}\,,\,x^{i}\right\}_{\NB}=0\,.
\ea
\label{conseq}
\end{equation}
In the following we consider the sector of solution-space with fixed on-shell
value of the Hamiltonian density
\begin{equation}
\cH_{\Mtwo}
=a^{-1}\sqrt{p^{2}+T^{2}a^{6}\det\cG}\equiv \Omega(\xi)\,.
\label{Omegadef}
\end{equation}
Since
$\Omega$ transforms as a scalar density,
fixing $\Omega$ finally  breaks the remaining static diffeomorphisms
down to the static area-preserving ones.
This gauge-fixed
sector is then equally described by a square-root free Lagrangian
\begin{equation}
\cL_{\Omega}:=
\half\left(\Omega a^{2}\,\dot{x}^{i}\dot{x}_{i}-\Omega^{-1}T^{2}
a^{4}\det\cG+\Omega\right)\,,
\label{LOmega}
\end{equation}
since the canonical momenta as well as the equations of motion are
identical to the on-shell ones of \eqref{LM2}.
Similarly to the   one-particle case in section~\ref{genconfmech},
the Hamiltonian constraint for $\cL_{\Omega}$ matches with \eqref{Omegadef} as
\begin{equation}
\cH_{\Omega}=\left(p^{2}+T^{2}a^{6}\det\cG-a^{2}\Omega^{2}\right)/
\left(2a^{2}\Omega\right)\equiv 0\,.
\label{HOmega}
\end{equation}
\indent
Identifying $\det\cG=m^{2}{/(T^{2}a^{4})}$ and
$\Omega=\mnut{/a^{2}(t)}$, the M2-brane Lagrangians (\ref{LM2}) and
(\ref{LOmega}) would respectively reduce to the point particle ones
(\ref{relatL}) and (\ref{nonrelLG}) for $\kappa=0$. However,
(\ref{conseq}) then  would imply    that $p^{2}-2m^{2}a^{2}$ is conserved and
the Hamiltonian constraint\,(\ref{HOmega}) could not be met.
Therefore, unlike in the flat background\,\cite{Banks:1996vh},
in an expanding Universe the M2-brane
dynamics cannot be consistently truncated to a point particle dynamics.\newline
\indent The matrix regularization of
 (\ref{LOmega}) prescribes to replace the dynamical fields
 $x^{i}(t,\xi^{\alpha})$ by time-dependent $N\times N$ Hermitian
matrices $X^{i}(t)$, the Nambu bracket
$\left\{x^{i},x^{j}\right\}_{\NB}$ by a matrix commutator
$i\!\left[X^{i},X^{j}\right]$\,\cite{deWit:1988ig},  and the worldvolume coordinates
$\xi^{\alpha}$ by non-dynamical matrices $\hat{\xi}^{\alpha}$
satisfying the non-commutative relation $[\hat{\xi}^{\alpha},\hat{\xi}^{\beta}]
=i\epsilon^{\alpha\beta}\times\mbox{constant}$\,\cite{Kim:2006wg}.  
The resulting   M2-brane matrix model   is of the general form
\begin{equation}
\displaystyle{\hat{\cL}_{\rm{M2}}\!=
\Tr\!\left[\frac{\,\hat{\Omega}a^{2}}{2}\!\left(\!D_{t}X^{i}\right)^{\!2}\!+
\frac{a^{4}}{l^{6}\hat{\Omega}}\!\left[X^{i},X^{j}\right]^{2}\!
+\frac{\hat{\Omega}}{2}\right],}
\label{LMMM2}
\end{equation}
where $\hat{\Omega}(t):=\Omega(t,\hat{\xi}^{\alpha})$ and $l$ is a
constant length.  The covariant time derivative
$D_{t}X^{i}=\dot{X}^{i}-i[A_{0},X^{i}]$ involves a non-dynamical gauge
field\,$A_{0}$, such that the matrix model admits a $\mbox{U}(N)$ gauge
symmetry.  With the canonical momenta $P^{i}=\half a^{2}
(\hat{\Omega}D_{t}X^{i}+D_{t}X^{i}\hat{\Omega})$,  ${\delta A}_{0}$
gives the Gauss constraint for the
 physical states $[P^{i},X_{i}]=0$, which is consistent with (\ref{conseq}).
Combining the  result of section~\ref{genconfmech}
and \eqref{LMMM2} with the identifications
$\hat{\Omega}(t)\equiv\mnut{/a^{2}}(t)$,  $\hat{\eta}\equiv\mnut$\,(\ref{mtdef}),
we finally   determine the precise form of  the matrix model for $N$ D-particles
in the flat RW background, for which the Hamiltonian is
\be
~\!\!\!\!\!\hat{\cH}_{{\rm D}0}\!=\!{\dis{\frac{1}{\,2\hat{\eta}(t)}}}
\Tr\!\left[P_{i}^{2}-2a^{6}(t){l^{-6}}\!\left[X^{i},X^{j}\right]^{2}
\!-\!(\nu m)^{2}\right].
\label{HMMD0}
\ee
One crucial difference between the two matrix models for the M2-brane\,\eqref{LMMM2}
and for D-particles\,\eqref{HMMD0} is
the last non-dynamical potential term which gives rise to a  different
Hamiltonian constraint, reflecting   the different dynamical
behavior of  M2-brane and D-particles. We interpret the
non-dynamical term as the temperature$~\cT$ of the Universe\,\cite{Weinberg}, since
from (\ref{HMMD0}) and $\hat{\cH}_{{\rm D}0}\equiv 0$
this term corresponds to the average  particle  energy.
For the D-particle case, we have
$\cT=m\nu^{2}/(2a\sqrt{a^{2}+\nu^{2}})$ which nicely interpolates between
the temperatures of the early radiation dominated  era,
$\cT\propto a^{-1}$ and the
late  matter dominated   era,
$\cT\propto a^{-2}$.\newline
\indent
Both in the Hamiltonian corresponding to (\ref{LMMM2}) and in  (\ref{HMMD0}),
the constraint~$\hat{\cH}{\equiv 0}$
gives rise to an explicit realization of  spacetime emergence:
${-\Tr[X^{i},X^{j}]^{2}=\Tr[X^{i},X^{j}][X_{i},X_{j}]^\dagger\geq 0}$ must
decrease on-shell as the scale factor $a(t)$ increases\,\footnote{
For (\ref{LMMM2})  note from (\ref{conseq}) $\,\partial_{t}(a^{4}{/\Omega^{2}})
\equiv 6a\dot{a}p^{2}/\Omega^{4}>0$.}.
Near the singularity  
the $X^{i}$ do not commute and thus cannot be simultaneously diagonalized.
This leaves the particle positions fuzzy, and hence the spacetime geometry they probe.
As the scale factor  grows or the Universe cools down,
the fuzziness  disappears and the classical
Robertson-Walker geometry emerges. In \cite{Craps}, a similar
scenario was found in plane wave backgrounds from the time-dependent coupling of the
Yang-Mills potential. Our matrix model is, however,  subject to the  additional
Hamiltonian constraint.\newline
\indent As a generalization of   the one-particle case in section~\ref{genconfmech},
the inhomogeneous fluctuations in the RW Universe are expected to be mapped to
the matrix model\,\eqref{LMMM2}.
Quantum corrections to our scenario can be calculated
in analogy to \cite{Craps}, in particular they may restrict  the
time dependence of $a(t)$.\newline
\indent We thank Evgeny Babichev, Ben Craps and Corneliu Sochichiu for helpful
discussions.  The research of JHP is supported by the Alexander von
Humboldt foundation and by Center for Quantum Spacetime of Sogang
University with grant number R11 - 2005 - 021.



\begin{thebibliography}{99}


\bibitem{Spergel:2006hy}
  D.~N.~Spergel {\it et al.},
``Wilkinson Microwave Anisotropy Probe (WMAP) three year results,"
astro-ph/0603449.




\bibitem{HashiSethi}
  A.~Hashimoto and S.~Sethi,
  Phys.\ Rev.\ Lett.\  {\bf 89}  261601 (2002).

\bibitem{Craps}
  B.~Craps, S.~Sethi and E.~P.~Verlinde,
  JHEP {\bf 0510}  005 (2005);
  M.~Li and W.~Song,
  JHEP {\bf 0608}, 089 (2006);
  B.~Craps, A.~Rajaraman and S.~Sethi,
  Phys.\ Rev.\ D {\bf 73} 106005 (2006);
  S.~R.~Das and J.~Michelson,
  Phys.\ Rev.\ D {\bf 73}  126006 (2006).


\bibitem{Banks:1996vh}
  T.~Banks, W.~Fischler, S.~H.~Shenker and L.~Susskind,
  Phys.\ Rev.\ D {\bf 55}  5112 (1997).

\bibitem{Maldacena:1997re}
  J.~M.~Maldacena,
  Adv.\ Theor.\ Math.\ Phys.\  {\bf 2}, 231 (1998)
  [Int.\ J.\ Theor.\ Phys.\  {\bf 38}, 1113 (1999)].
\bibitem{Weinberg}
 S.~Weinberg,
 ``Gravitation and Cosmology,"
  John Wiley \& Sons, Inc. (1972).


\bibitem{deAlfaro:1976je}
  V.~de Alfaro, S.~Fubini and G.~Furlan,
  %
  Nuovo Cim.\ A {\bf 34}, 569 (1976).





\bibitem{Erdmenger:2006eh}
J.~Erdmenger, J.-H.~Park and C.~Sochichiu,
Class.\ Quant.\ Grav.\  {\bf 23} 6873 (2006).


%

\bibitem{Pioline:2002qz}
  B.~Pioline and A.~Waldron,
  Phys.\ Rev.\ Lett.\  {\bf 90} 031302 (2003).



\bibitem{Belinsky}
  V.~Belinsky, I.~Khalatnikov and E.~Lifshitz,
Adv.\ Phys.\  {\bf 19}, 525 (1970);  Adv.\ Phys.\  {\bf 31}, 639 (1982).


\bibitem{Witten:1995im}
  E.~Witten,
  Nucl.\ Phys.\ B {\bf 460} 335 (1996).


\bibitem{Kim:2006wg}
  N.~Kim and J.-H.~Park,
  Nucl.\ Phys.\ B {\bf 759} 249 (2006).


\bibitem{deWit:1988ig}
  B.~de Wit, J.~Hoppe and H.~Nicolai,
  Nucl.\ Phys.\ B {\bf 305} 545 (1988).








\end{thebibliography}

\end{document}